\documentclass[twocolumn,showpacs,preprintnumbers,amsmath,amssymb]{revtex4}

\usepackage{graphicx}
\usepackage{rotating}
\usepackage{dcolumn}
\usepackage{bm}

\begin{document}


\title{Nonsensitive nonlinear homotopy approach}
\author{Yuan Gao$^{1}$ and S. Y. Lou$^{1,2,3}$}
\affiliation{$^{1}$Department of Physics, Shanghai Jiao Tong
University, Shanghai, 200240, China\\
$^{2}$Faculty of Science, Ningbo University, Ningbo, 315211,
China\\
$^3$School of Mathematics, Fudan University, Shanghai, 200433,
China}

\date{\today}

\begin{abstract}
Generally, natural scientific problems are so complicated that one
has to establish some effective perturbation or nonperturbation
theories with respect to some associated ideal models. In this
Letter, a new theory that combines perturbation and nonperturbation
is constructed. An artificial nonlinear homotopy parameter plays the
role of a perturbation parameter, while other artificial nonlinear
parameters, of which the original problems are independent,
introduced in the nonlinear homotopy models are nonperturbatively
determined by means of a principle minimal sensitivity. The method
is demonstrated through several quantum anharmonic oscillators and a
non-hermitian parity-time symmetric Hamiltonian system. In fact, the
framework of the theory is rather general that can be applied to a
broad range of natural phenomena. Possible applications to condensed
matter physics, matter wave systems, and nonlinear optics are
briefly discussed.
\end{abstract}


\maketitle
In general, natural scientific problems cannot be exactly described
by some analytical expressions. What scientists usually do is just
first to establish some ideal models, and then to consider
additional effects by use of some perturbation and nonperturbation
theories.

Perturbation theories (PTs) \cite{YangJK,Ablowitz} already have
quite a long and interesting history in physics. They have been used
to explore physical systems that can not be solved exactly but with
a small parameter. Though great challenges are coming from the
blooming numerical calculation methods, PTs are still showing their
great power in solving various problems. For instance, perturbation
theories have been used to study the spectrum of non-hermitian
parity-time ($\mathcal{PT}$) symmetric Hamiltonian \cite{PTS}, the
structure of trapped Bose-Einstein condensates (BECs) with
long-range anisotropic dipolar interactions \cite{dBEC}, and the
interacting fermions in two-dimensions \cite{Fermion}. Besides, the
fundamental ideas of PTs have been utilized to study some models in
quantum computation theory (QCT). Techniques based on PTs have also
been applied to engineer interesting Hamiltonians, whereby the
outstanding perturbation theory gadgets (PTGs) technique was
developed \cite{MB1,MB2}. A short but incisive review of the history
of PT and its new development in quantum many-body theory, quantum
computation and quantum complexity theory has been given recently by
M. M. Wolf \cite{Wolf}.

On the other hand, nonperturbation theories (NPTs) have been
established to solve real scientific problems with the lack of a
small parameter that acts as a perturbation parameter. In this case,
some parameters will be artificially introduced as some formal
perturbation parameters and/or nonperturbation parameters. Usually,
one of these artificial perturbation parameters have to be set
finite, say, the unit one, while others will be determined in some
nonperturbative ways such as some types of optimized approaches.

Considering the fact that physical quantities should be independent
of any particular perturbation and/or nonperturbation method used to
calculate them, Stevenson \cite{1981} proposed an optimized
perturbation theory (OPT) with the principle of minimal sensitivity
(PMS) as a key ingredient. Some other perturbation methods appeared
in 1970's and 1980's, such as the non-parameter expansion method
\cite{1979}, and the linear $\delta$ expansion method (LDE)
\cite{8788,1990,LuWF,2005,1993,1993a,1995}, have the same idea as
implied in the OPT. These methods, especially the LDE, have found
successful applications in many physical contexts during the past
three decades.

Actually, as we will show later in this Letter, the basic idea
underlying those methods mentioned above can be integrated into a
general framework of the (linear) homotopy analysis method (HAM)
\cite{kowalski,liao}. In the language of field theory, it is
possible to construct a new action $S_q=\mathcal{H}(S_0,S,q)$ by
building a homotopy relation between an original action $S$ and a
trial ideal action $S_0$ (which is solvable and reflects as much as
possible physics of $S$). Here $S$ can represent any object or
quantity for any scientific problem, such as the Green function,
density operator, distribution function, generating functional,
differential equations, and so on. $q\in [0,1]$ is called a homotopy
parameter. $S_q$ is fixed at the two endpoints, i.e., $S_{q=0}=S_0$
and $S_{q=1}=S$, while can possess different expressions when $q\neq
0,1$, depending on $\mathcal{H}(S_0,S,q)$. Hence, different homotopy
relations yield different actions. For example, a linear homotopy
relation yields an action as $S_q=(1-q)S_0+qS$, which is exactly the
form used in the LDE and other methods mentioned above. However, the
homotopy relation is by no means constrained to be linear. We can
introduce a simple and direct nonlinear extension
\begin{eqnarray}
S_q&=&(1-q)S_0+\sum_{i=1}^nq^iS_i+q^{n+1}\left(S-\sum_{i=1}^nS_i\right).\label{HR}
\end{eqnarray}
In the standard procedure of the LDE, $S_0$ contains one or more
auxiliary parameters $\eta_i$. In our generalized nonlinear homotopy
relation \eqref{HR}, auxiliary actions $S_1,S_2,\ldots,S_n$ also
contain some parameters $\xi_i$. These parameters are also
artificial, which means that $S_{q=1}=S$ is independent of both
$\eta_i$ and $\xi_i$. This can be easily verified with the help of
Eq. \eqref{HR}, as all $S_i~(i=1\ldots n)$ will be eliminated when
the nonlinear homotopy parameter $q$ is set to 1. In order to
describe it more explicit and without losing the nonlinear property
of Eq. \eqref{HR}, let us throw away $O(q^3)$ terms to obtain
\begin{eqnarray}
S_q&=&(1-q)S_0(\eta)+qS_1(\xi)+q^{2}(S-S_1(\xi)),\label{HR2}
\end{eqnarray}
with only two auxiliary parameters $\eta$ and $\xi$. Any desired
physical quantity $\Phi$ can be evaluated as a perturbation series
of $q$, which is set equal to 1 at the final step. If the series is
truncated at $q^N, N\geq n+1=2$, the approximant $\Phi^{(N)}$ is
obviously dependent of $\eta$ and $\xi$, while it will have no
relation with those two auxiliary parameters when
$N\rightarrow\infty$. In practice, $N$ is always finite, it is
therefore necessary to choose these parameters nonperturbatively to
make the expansion reasonable. We will take the PMS as an effective
criterion. According to the PMS, the auxiliary parameters $\eta$ and
$\xi$ should be selected to minimize the sensitivity of the
perturbation expansion to some small variations in them. It is
remarkable that a nonlinear homotopy relation and the combination of
the perturbation and nonperturbation theories via the PMS are two
fundamental ingredients of our nonsensitive nonlinear homotopy
approach (NNHA).

Before moving to concrete examples, let us discuss a little more
about the PMS. There is a natural way to fix the parameters $\xi$
and $\eta$, which is to solve the following equation system
\begin{eqnarray}
\frac{\partial\Phi^{(N)}(\eta,\xi)}{\partial\eta}=0,\quad
\frac{\partial\Phi^{(N)}(\eta,\xi)}{\partial\xi}=0 \label{3}
\end{eqnarray}
similar to the usual OPT theory. However, the above system does not
always have one solution. Eqs. \eqref{3} might have many solutions
$\{\xi_i, \eta_i\}$, therefore, we have to find a method to fix one
of them according to the PMS. To this end, we define the $M$-th
sensitivity $\kappa^{(N)}_M$
\begin{eqnarray}
\kappa^{(N)}_M=\sum_{j=1}^M\left\|\frac{\textrm{d}^M\Phi^{(N)}}
{\textrm{d}\xi^j\textrm{d}\eta^{M-j}}\right\|,\label{4}
\end{eqnarray}
where $\|\cdot \|$ denotes the norm. It is clear that the
sensitivities $\kappa^{(N)}_M$ for all $M$ should tend to zero as
$N\rightarrow \infty$, because the real model is $\{\xi, \eta\}$
independent. Consequently, it is natural to pick out the final
required $\{\xi,\eta\}$ by minimizing the first few sensitivities
$\kappa^{(N)}_i (i=1, 2, ...,k)$ for the smallest $k$. More details
can be seen in the following concrete examples. It is clear that a
solution (if it exists) of Eqs. \eqref{3} is really a minimum of the
first sensitivity (FS) $\kappa^{(N)}_1$.

Now we use our theory to study an eigenvalue problem of a given
Hamiltonian. This is a basic problem included in almost all quantum
physical problems. For instance, various models in condensed matter
physics deal with ground-states of spin Hamiltonians, and quantum
computation can be performed by encoding a solution of a computation
problem into the ground-state of a Hamiltonian \cite{Wolf}. The
calculation of energy levels for a Hamiltonian system is also a
crucial problem in solid physics \cite{Fermion}. As the first
example, we apply the theory to calculate the ground-state energy of
a sextic quantum anharmonic oscillator
\begin{eqnarray}
H=p^2+x^2+x^6. \label{sextic}
\end{eqnarray}
Generally, through a nonlinear homotopy relation in the form of Eq.
\eqref{HR2}, we can construct a new Hamiltonian
\begin{eqnarray}
H_q=(1-q)(H_0)+q(H_1)+q^2(H-H_1).\label{H2}
\end{eqnarray}
Here we choose $H_0=p^2+\omega x^2$ and $H_1=p^2+x^2+\lambda x^6$,
and thus obtain
\begin{eqnarray}
H(q,\omega,\lambda)&=&(1-q)(p^2+\omega
x^2)+q(p^2+x^2+\lambda x^6)\nonumber\\
&&+q^2(1-\lambda)x^6. \label{sextic}
\end{eqnarray}
It is obvious that $H_0$ is the Hamiltonian of a harmonic oscillator
with an auxiliary parameter $\omega$ denoting the frequency of the
oscillation. $H_1$ is a new term which can not be included without
the second order term of $q$. It is an auxiliary Hamiltonian with a
parameter $\lambda$ denoting the intensity of an additional sextic
effect. It is easy to verify that $H(1,\omega,\alpha)=H$, and
$H(0,\omega,\alpha)=H_0$.

Making advantage of Bender and Wu's work \cite{Wu}, we obtain the
ground-state energy of $H(q,\omega,\alpha)$ as a power series of
$q$,
\begin{eqnarray}
E_{HT}(q,\omega,\lambda)=\omega-\frac{(15\lambda-4\omega^4+4\omega^2)}{8\omega^3}q+\ldots,
\end{eqnarray}
which can be easily calculated to any order of $q$, with the help of
the symbolic calculation softwares. Using MAPLE, we calculate the
series up to order 21, and obtain an analytical expression of the
approximate ground-state energy $E^{(21)}_{HT}(q,\omega,\lambda)$,
and finally set $q=1$.

Now comes the problem to determine parameters $\omega$ and $\lambda$
under the PMS. Roughly speaking, the PMS tends to pick out
parameters around which the profile of
$E^{(21)}_{HT}(1,\omega,\lambda)$ is the ``flattest". So what can be
learned from the profile of $E^{(21)}_{HT}$ as displayed Fig. 1,
where the symbolic $E$ denotes $E^{(21)}_{HT}$? It is discovered
that when $(\omega,\lambda)$ varies in the region
$[3.7,4]\times[0.25,0.4]$, the variation in the energy
$E^{(21)}_{HT}$ is less than $10^{-6}$. We take a very accurate
value $E_{ex}=1.435\ 624\ 619$ as an ``exact" result, which is given
by Mei{\ss}ner and Steinborn \cite{exact}. Thus we have an error
$E^{(21)}_{HT}-E_{ex}$ vary from $-4.62\times10^{-6}$ to
$5.38\times10^{-6}$ in the above region. In the case that the exact
value of a solution is not known, we can calculate
$E^{(21)}_{HT}-E^{(20)}_{HT}$ as an error. It is reasonably found
that $E^{(21)}_{HT}-E^{(20)}_{HT}\sim 10^{-6}$. To pick out an
optimal point in this region, let us go further from the figure to
determine a pair of nonsensitive parameters numerically. In Fig. 2
we draw the curves of $\frac{\partial
E^{(21)}_{HT}}{\partial\omega}=0$ (solid) and $\frac{\partial
E^{(21)}_{HT}}{\partial\lambda}=0$ (dash), and find that they do not
intersect in this region. In this situation, we can not obtain a
pair of $(\omega, \lambda)$ to vanish these two first-order partial
derivatives simultaneously. Motivated by the spirit of the PMS, we
minimize the FS for $\Phi^{(N)}=E^{(21)}_{HT}(\omega,\lambda)$ by
simply taking the norm as $\|A\|=A^2$. Consequently, a pair of
parameters $\omega=3.847\ 166$ and $\lambda=0.377\ 585$ is obtained
to minimize the FS. The substitution of these parameters into the
approximant $E^{(21)}_{HT}(1,\omega,\lambda)$ leads to an
approximate energy $E^{(21)}_{HT}=1.435\ 624$ with an error of
$-5.53\times10^{-7}$.

\input epsf
\begin{figure}
\epsfxsize=4.5cm\epsfysize=3cm\epsfbox{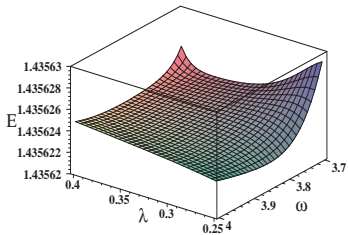}
\caption{$E^{(21)}_{HT}$ as a function of $\omega$ and $\lambda$ in
the region $[3.7,4]\times[0.25,0.4]$.}
\end{figure}

\input epsf
\begin{figure}
\epsfxsize=4.5cm\epsfysize=3cm\epsfbox{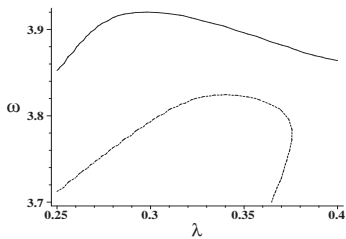} \caption{The curves
of $\frac{\partial E^{(21)}_{HT}}{\partial\omega}=0$ (solid) and
$\frac{\partial E^{(21)}_{HT}}{\partial\lambda}=0$ (dash). }
\end{figure}

As a compare, we also give an approximant truncated at order 21 for
the ground-state energy by using a linear homotopy relation as
$H(q,\omega)=(1-q)(p^2+\omega x^2)+q(p^2+x^2+x^6)$, which has been
applied in the LDE method. It yields an approximant
$E^{(21)}_{LDE}(q,\omega)$ which only depends on $\omega$, after $q$
is set to be 1. After some similar procedures, we obtain a final
result $E^{(21)}_{LDE}(1,\omega_0)=1.435\ 428$ with $\omega_0=5.254\
070$ satisfying ${\textrm{d}E^{(N)}}/{\textrm{d}\omega}=0$. The
error of $E^{(21)}_{LDE}$ is $-1.97\times10^{-4}$, which is about
$10^3$ times larger than that of $E^{(21)}_{HT}$. The above example
and the corresponding numerical results show that our theory does
largely improve the accuracy of the approximant without more
difficulty. We have also study the ground and excited states for
some quartic and octic anharmonic oscillators in the same way, and
drawn the same conclusions. It is remarkable that when the
coefficients of the anharmonic terms are large, the method is still
applicable and the accuracy is at the same level.

\input epsf
\begin{figure}
\epsfxsize=4.5cm\epsfysize=3cm\epsfbox{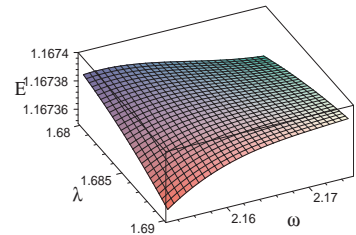}
\caption{$E^{\prime(21)}_{HT}$ as a function of $\omega$ and
$\lambda$ in the region $[1.68,1.69]\times[2.153,2.175]$.}
\end{figure}

Next we go further to study non-hermitian Hamiltonian systems.
Recently, many progresses have been made for non-hermitian
$\mathcal{PT}$-symmetric physics. We are thus motivated to
investigate the energy spectrum of a non-hermitian Hamiltonian with
$\mathcal{PT}$-symmetry. Bender and his co-workers have demonstrated
that such a $\mathcal{PT}$-symmetric Hamiltonian
\begin{eqnarray}
H^\prime=p^2+\frac14x^2+ix^3,
\end{eqnarray}
has a real and positive spectrum \cite{Bender}. According to our
theory, the nonlinear homotopy relation can also be built to link a
non-hermitian Hamiltonian with some auxiliary Hamiltonians that can
be either hermitian or non-hermitian. For instance, a nonlinear
homotopy Hamiltonian can be constructed as
\begin{eqnarray}
H^\prime(q,\omega,\lambda)&=&(1-q)(p^2+\omega
x^2)+q(p^2+\frac14x^2+i\lambda x^3)\nonumber\\
&&+iq^2(1-\lambda)x^3. \label{sextic}
\end{eqnarray}
It can be seen that $H^\prime(q,\omega,\lambda)$ also has the
$\mathcal{PT}$-symmetry, and its ground-state eigenvalue can be
expressed as
\begin{eqnarray}
E^\prime_{HT}(q,\omega,\lambda)=\omega+\frac{1-4\omega^2}{8\omega}q+\ldots,
\end{eqnarray}
where the coefficients of each order of $q$ are real. Truncating the
series at order 21 and setting $q=1$ yield an approximant
$E^{\prime(21)}_{HT}$ for the ground-state energy of $H^\prime$. The
pair that eliminates the FS \eqref{4} for
$\Phi^{(N)}=E^{\prime(21)}_{HT}(\omega,\lambda)$ is $\{\omega=2.160\
887,\lambda=1.684\ 022\}$, and the profile of $E^{\prime(21)}_{HT}$
around this stationary point is shown in Fig. 3 with $E$ denoting
$E^{\prime(21)}_{HT}$. At the stationary point,
$E^{\prime(21)}_{HT}$ equals $1.167\ 384$, with an error
$7.61\times10^{-5}$ compared with the numerical result $1.167\ 46$
given by Bender and Dunne \cite{JMP}.

This example reveals great validity of our theory on investigating
non-hermitian $\mathcal{PT}$-symmetric Hamiltonian systems. Besides,
our theory can be used to study more complicated problems beyond a
linear eigenvalue problem. It is shown in a series of recent works
that nonlinear localized structures can be generated in
$\mathcal{PT}$ periodic potentials \cite{soliton-pt,beam}. Optical
solitons are produced by adding a self-focusing Kerr nonlinear
$\mathcal{PT}$-symmetric potential $iW(x)$, which satisfies
$W(-x)=-W(x)$, to the usual nonlinear Schr\"odinger equation. Our
theory can also be applied to investigate this type of nonlinear
$\mathcal{PT}$-symmetric system by building suitable homotopy
relations associating a system hard to be exactly solved with those
solvable. In this way, the knowledge of the latter can be used to
explore the original system approximately, and the necessary
accuracy is entirely ensured by the principle of minimal
sensitivity.

As is known, physical models that can be solved exactly are very
rare. In a recent work on the structure of trapped Bose-Einstein
condensates (BECs) with long-range anisotropic dipolar interactions,
it is found that a small perturbation in the trapping potential can
lead to dramatic changes in the condensate's density profile for
sufficiently large dipolar interaction strengths and trap aspect
ratios\cite{dBEC}. The method the authors used is the traditional
perturbation method which require the perturbation to be small, our
theory can thus be used to extend their work to the range where the
perturbation is larger. So matter-wave systems with dipolar
interactions would be a suitable field for our theory. Besides, the
theory can also facilitate the study of quantum many-body systems,
where perturbation techniques are essential tools\cite{Wolf}.

To conclude, we have proposed a novel approach called NNHA, a
combination theory of perturbation and nonperturbation with the help
of the PMS and nonlinear homotopy realization. The new method relies
on a nonlinear homotopy to construct an approximant with auxiliary
parameters that does not exist in the original models. The nonlinear
homotopy parameter $q$ is taken as a perturbation parameter and
finally fixed to 1 as usual, because the original model is related
to $q=1$ only. Other parameters are nonperturbatively fixed at the
end of the calculation by implementing the PMS, which requires the
approximant have the least dependence on these auxiliary parameters
for the reason that the original system is independent of them. The
theory has been applied to study the energy spectrum of some
hermitian Hamiltonian systems as well as non-hermitian
$\mathcal{PT}$-symmetric Hamiltonians. Highly accurate numerical
results demonstrate the validity of the theory. Possible
applications can be made further to $\mathcal{PT}$-symmetric optical
systems, dipolar Bose-Einstein condensates (BECs), and quantum
many-body systems. Actually, we do believe that the method can be
used to solve any scientific problem where mathematics is required.

The authors are indebt to thank Dr. X. Y. Tang, C. L. Chen, and M.
Jia for their helpful discussions. The work was sponsored by the
National Natural Science Foundation of China (Nos. 10735030 and
90503006), the National Basic Research Program of China (973 Program
2007CB814800) and the PCSIRT (IRT0734).

\end{document}